\documentclass[peerreview,draftcls,print]{ieeecolor2}
\usepackage{generic}
\usepackage{cite}
\usepackage{amsmath,amssymb,amsfonts}
\usepackage{algorithmic}
\usepackage{graphicx}
\usepackage{textcomp}
\usepackage{subcaption}
\usepackage{booktabs} 
\usepackage{multirow}
\def\BibTeX{{\rm B\kern-.05em{\sc i\kern-.025em b}\kern-.08em
    T\kern-.1667em\lower.7ex\hbox{E}\kern-.125emX}}
\begin{document}
\title{A High-Fidelity Neurosurgical Training Platform for Bimanual Procedures: A Feasibility Study}
\author{Houssem-Eddine Gueziri, \and Abicumaran Uthamacumaran, \and Widad Safih, \and Abdulrahman Almansouri, \and Nour Abou Hamdan, \and José A. Correa, \and Étienne Léger, \and D. Louis Collins, \and Rolando F. Del Maestro
\thanks{This work was supported by grants from the Canadian Institutes of Health Research (195899), Brain Tumour Foundation of Canada-Brain Tumour Research Grant.}
\thanks{H.-E. Gueziri is with the TÉLUQ University, Montreal, QC H2S 3L5 Canada. (e-mail: houssem.gueziri@teluq.ca).}
\thanks{A. Uthamacumaran, W. Safih, A. Almansouri, N. Abou Hamdan, Rolando F Del Maestro, are with the Neurosurgical Simulation and Artificial Intelligence Learning Centre, Montreal, QC H2X 4B3 Canada.}
\thanks{J. A. Correa, É. Léger, D. L. Collins, are with McGill University, Montreal, QC H2X 4B3 Canada.}}

\maketitle

\begin{abstract}

\textbf{Background}. Bimanual psychomotor proficiency is fundamental to neurosurgical procedures, yet it remains difficult for trainees to acquire and for educators to objectively evaluate performance. 
In this study, we investigate the feasibility of a neurosurgical simulation platform that integrates an anatomically realistic brain model with surgical instrument tracking to support training and objective assessment of bimanual tasks in the context of subpial corticectomy.

\textbf{Methods}. 
We developed and evaluated a neurosurgical simulation platform based on an \textit{ex-vivo} calf brain model
and a multi-camera tracking system capable of simultaneously capturing the motion of surgical instruments in both hands, including collection of real-time instrument trajectories and synchronized video recordings.
These enabled extraction of motion-based, time-based, and bimanual coordination metrics.
We conducted a case series involving 47 participants across four training levels: medical students, junior residents, senior residents, and neurosurgeons.

\textbf{Results}. The tracking system successfully captured instrument motion during 81\% of the periods when instruments were actively used throughout the simulation procedure.
Several extracted metrics were able to significantly differentiate between levels of surgical expertise. 
In particular, instrument usage duration and custom-defined bimanual coordination metrics such as instrument tip separation distance and simultaneous usage time, show potential as features to identify participant expertise levels with different instruments.

\textbf{Conclusions}. We demonstrated the feasibility of tracking surgical instruments during complex bimanual tasks in an ex-vivo brain simulation platform. The metrics developed provide a foundation for objective performance assessment and highlight the potential of motion analysis to improve neurosurgical training and evaluation.
\end{abstract}


\section{Background}
\label{sec:background}
\IEEEPARstart{E}{pilepsy} is a neurological disorder characterized by recurrent seizures that affects 300,000 Canadians and more than 70M individuals around the world\cite{Fisher2014}. 
Patients with epilepsy see their quality of life significantly affected due to uncontrollable seizures that lead to more frequent fractures and bruises due to injuries; in addition to psychological conditions such as anxiety and depression\cite{who2023report}. 
Mortality rates among people with epilepsy are three times the rate of the general population, and sudden death rates are more than twenty times higher\cite{Hitiris2007}. 
While medication is the primary treatment for epilepsy, 20-40\% of patients exhibit drug-resistant epilepsy\cite{Vaughan2018}, in which case surgical treatment is considered. 
Subpial corticectomy is a surgical procedure that involves removing pathological lesions responsible for focal seizures while preserving the pia and minimizing damage to the surrounding eloquent tissue. 
Acquiring proficiency in the corticectomy technique is crucial for neurosurgery trainees in both epilepsy and tumor resection procedures, as research has established a strong correlation between surgeon skill in performing bimanual psychomotor tasks and patient outcomes\cite{Birkmeyer2013,Brightwell2013}. 
This performance is often related to minimizing or preventing surgical errors. 
When examining all surgeries, Baker et al. reported that 185,000 patients out of 2.5 million admissions per year in Canadian hospitals were exposed to a surgical adverse event and 36\% of these events were considered preventable\cite{Baker2004}. 
In a report published by the Institute of Medicine (US) Committee on Quality of Health Care in America\cite{Kohn2000}, the authors conclude that many surgical errors can be attributed to suboptimal surgical training. 
In addition, protocols restricting duty hours have been implemented as a strategy to reduce the risk of fatigue in resident work schedules\cite{Johns2009} and other restrictions due to the COVID-19 pandemic have decreased resident operating room training, limiting opportunities to acquire and refine essential bimanual surgical skills.

Surgical training is evolving from an apprenticeship model to more competency-based educational frameworks\cite{Franzese2007,VanHeest2022}.
These frameworks need to have assessment capability based on quantifiable objective metrics that are transparent to both the educator and the trainee\cite{Mirchi2020,Yilmaz2022}.
However, the existing evaluation process for assessing the psychomotor skills of surgical residents relies heavily on subjective judgments made by consultant surgeons. 
This method is intrinsically incomplete as it does not capture the full spectrum of competencies necessary for modern surgical proficiency.
Surgical simulators have been used to create safe learning and training environments by simulating complex patient operative pathologies in risk-free environments\cite{Coelho2014}.
Yet, neurosurgical simulation training is not currently integrated into core curriculums due to several factors.
While virtual and augmented reality simulators enable a variety of cases and provide metrics that track the quality of performance of the specific task, they often lack realism.
Although commercial systems, such as ImmersiveTouch\cite{Luciano,Alaraj2013} and NeuroTouch\cite{Delorme2012} provide haptic feedback for more immersion, tactile feedback can be limited and the virtual rendering makes it feel like a “video arcade machine”\cite{Cobb2016}.
Furthermore, the high costs associated with these systems restrict their availability in centers with limited funding.
Live animal brains have also been employed in neurosurgical simulation for aneurysm clipping\cite{Olabe2009}, microvascular anastomosis\cite{Schoffl2006,Colpan2008,Olijnyk2019}, and tumor resection\cite{Kamp2015,Altun2019,Valli2019}.
While they provide high tissue fidelity, most platforms based on animal brains lack the ability to capture objective measures of surgical performance.
In addition, ethical considerations when using live animals for experimentation need to be considered, and animal care may increase the costs. 

In our previous work, we developed an initial prototype of a simulation platform that combines an ex-vivo calf brain, to reproduce realistic bimanual corticectomy scenarios, with an optical tracking system to collect real-time instrument trajectory data\cite{Winkler2020}. 
Bovine and calf brains have been widely used in neurosurgical training due to their close mechanical and anatomical resemblance to human brain tissue, their availability from local food suppliers, and the minimal ethical concerns associated with their use\cite{Alsayegh2021}. 
In a recent study we demonstrated both \textit{face validity} (i.e., the extent to which assessment conditions resemble a real-life situation) and \textit{content validity} (i.e., the extent to which an assessment tool measures the intended attributes accurately)\cite{Almansouri2022}.
In this paper, we explore the feasibility of the simulation platform to capture surgical performance across different levels of expertise. 
While this study focuses on the technical design, the quality of the captured data, and the characterization of the surgical task, the assessment of surgical performance using these metrics remains beyond its scope. 
Our objective is to present the capabilities of the platform and provide preliminary evidence supporting its potential utility in neurosurgical training.


\section{Related work}
\label{sec:related_work}

Simulation-based training has been shown to significantly improve procedural knowledge and technical proficiency in various surgical disciplines. 
In neurosurgery, there is an increasing willingness to integrate simulation platforms into training programs\cite{Davids2021}.
Current simulation modalities include high-fidelity physical models, virtual and augmented reality environments, and 3D-printed anatomical replicas designed to closely mimic real-world operative conditions.
These technologies are increasingly used in subspecialties such as neurovascular and skull base surgery, providing safe, repeatable training environments \cite{dadario2021}.

High-fidelity, physically immersive simulation platforms employ anatomically realistic models that replicate the tactile and visual characteristics of live surgery, enabling trainees to practice complex procedures in a controlled, reproducible environment. 
Such models have been developed for a variety of surgical applications, including palatoplasty\cite{Ghanem2019} and cheiloplasty\cite{Podolsky2018}, providing structured environments for cleft repair training,  
cerebrovascular bypass\cite{Cikla2018}, supporting microsurgical skill acquisition in neurovascular anastomosis, and transorbital ultrasound measurement\cite{Hajat2020}.
As a result, such models are progressively replacing traditional cadaver training methods, offering enhanced accessibility, repeatability, and the ability to simulate a broader range of pathological scenarios. 

Another alternative is 3D printing, which enables the creation of patient-specific anatomical models using high-resolution imaging data, such as CT or MRI scans.
These models provide a cost-effective and customizable platform for rehearsing complex neurosurgical procedures, including tumor resections and vascular interventions. 3D printed simulators offer visual and tactile realism, support preoperative planning, and facilitate hands-on skill acquisition. However, limitations include the lack of soft tissue fidelity, the static nature of printed models, and the inability to simulate dynamic physiological responses such as bleeding or tissue deformation.

Augmented reality, virtual reality, and mixed reality simulator platforms have also emerged as valuable tools in neurosurgical education, offering computer-generated interactive environments that enhance the learning experience. 
In augmented reality, digital content is projected onto the real world, enabling users to interact with virtual anatomical structures while maintaining awareness of their surroundings.
For example, augmented reality simulation has been employed in ventriculostomy\cite{yudkowsky2013practice}.
In contrast, virtual reality fully immerses users in a 3D synthetic environment to interact with simulated surgeries, and has been used, for example, in cranial neurosurgery\cite{gasco2013novel}, aneurysm clipping \cite{gmeiner2018virtual}, and tumor resection \cite{sawaya2018virtual}.
Finally, mixed reality integrates the components of both augmented and virtual reality, facilitating the interaction between tangible and digital entities through the incorporation of spatial and gesture tracking technologies, such as in ventriculostomy\cite{hooten2014mixed} or craniectomy \cite{coelho2019development}.

Commercial platforms like ImmersiveTouch (Immersivetouch Inc., Westmont, IL, USA) \cite{Luciano,Alaraj2013}, mainly used for cranial and spinal procedures, offer virtual reality-based simulation with stereoscopic visualization and haptic feedback; ANGIO Mentor (3D Systems, Littleton, CO) focuses on endovascular interventions that simulate catheter-based procedures in a high-fidelity vascular environment; NeuroSIM (VRmagic GmbH, Mannheim, Germany) \cite{beier2011neurosim} was designed for training in cranial micro-neurosurgery, emphasizing the development of psychomotor skills through task-based scenarios and force-feedback instrumentation; and NeuroVR
(CAE Healthcare Inc., Saint-Laurent, QC, Canada) offers virtual reality-based simulation for neurosurgical training combining realistic 3D models, tactile feedback and performance metrics.

Despite their usefulness, these technologies have several limitations. Virtual and mixed reality systems often lack the tactile fidelity required for precise neurosurgical training, making it difficult to replicate the nuanced feel of tissue manipulation. 
In addition, high development and equipment costs can limit accessibility, especially in resource-constrained settings.
AR systems, while offering contextual awareness, may suffer from limited field of view, registration errors, or hardware bulkiness, which can interfere with task execution.
Furthermore, while many platforms provide metrics on user interaction, the clinical validity and standardization of these assessments remain underdeveloped.
Our platform combines an ex-vivo calf brain model to replicate anatomical and tactile realism and a multi-camera tracking system to capture objective metrics of surgical performance \cite{Winkler2020,Almansouri2022,tran2021}.

\section{Methods}
\label{sec:methods}

\subsection{Surgical procedure}
\label{subsec:procedure}

Corticectomy is used predominantly to treat drug-resistant focal epilepsy by excising the epileptogenic cortex, which is the region of the brain responsible for the generation of seizures, while maintaining the integrity of neighboring functional areas.
In our procedure, the subpial corticectomy was performed using a pair of \textit{microscissors} to make the initial incision in the pia mater, followed by the use of \textit{bipolar forceps} to elevate the pia, and a \textit{SONOPET ultrasonic aspirator} (Stryker, Kalamazoo, MI, USA) to remove designated cortical regions. 
The procedure was performed under an OPMI Pico surgical microscope (Carl Zeiss Meditec AG, Oberkochen, Germany), replicating the operative conditions.
A Blackfly S GigE camera (FLIR, Wilsonville, OR, USA) was mounted into the microscope's optical path to capture video recordings of the procedure for further analysis. 
It is important to note that this camera was used solely for video recording and did not contribute to instrument trajectory tracking.

\subsection{Study design}
\label{subsec:study}


\begin{table}[b]
    \caption{Distribution of participants according to their training level and expertise. PGY: Post-Graduate Year.}
    \centering
    \begin{tabular}{llllcc}
        \toprule
         & & \multicolumn{4}{c}{\textbf{Groups}}\\
         \cline{3-6}
         & & \multicolumn{1}{c}{\textbf{Student}} & \multicolumn{1}{c}{\textbf{Junior}} & \multicolumn{1}{c}{\textbf{Senior}} & \multicolumn{1}{c}{\textbf{Expert}} \\
        \midrule
        \multicolumn{2}{l}{\textbf{Training level}} & & & & \\
        & Medical student & 14 (30\%) &  &  &  \\
        & Resident PGY 1 &  & 5 (11\%) &  &  \\
        & Resident PGY 2 &  & 3 (6\%) &  &  \\
        & Resident PGY 3 &  & 2 (4\%) &  &  \\
        & Resident PGY 4 &  & 1 (2\%) &  &  \\
        & Resident PGY 5 &  & 3 (6\%) &  &  \\
        & Resident PGY 6 &  &  & 4 (8\%) &  \\
        & Fellow Spine &  &  & 2 (4\%) &  \\
        & Fellow Epilepsy &  &  & 1 (2\%) &  \\
        & Fellow Pediatrics &  &  & 2 (4\%) &  \\
        & Fellow Oncology &  &  & 1 (2\%) &  \\
        & Fellow Functional &  &  & 1 (2\%) &  \\
        & Neurosurgeons &  &  &  & 8 (17\%) \\
        \midrule
        \multicolumn{2}{l}{\textbf{Dominant hand}} &  &  & &  \\
        & Right & 13 & 12 & 9 & 8 \\
        & Left & 1 & 2 & 2 & 0 \\
        \midrule
        \multicolumn{2}{l}{\textbf{Total}} & 14 (30\%) & 14 (30\%) &  11 (23\%) & 8 (17\%)  \\
        \bottomrule
    \end{tabular}
    \label{tab:groups}
\end{table}

To assess the technical feasibility and data quality of the simulation platform, we conducted a case series that involved participants from a spectrum of surgical training backgrounds. 
Participants were categorized into four distinct groups based on their clinical experience and training level: \textit{Student}, \textit{Junior}, \textit{Senior}, and \textit{Expert}, as detailed in Table\,\ref{tab:groups}. 
The \textit{Expert} group comprised eight fully neurosurgeons with independent operative experience. 
The \textit{Senior} group included seven post-graduate year 6 (PGY-6) residents and clinical fellows specialized in epilepsy, neurosurgical oncology, pediatric neurosurgery and other neurosurgical subspecialties.
The \textit{Junior} group was composed of 14 residents in the early stages of their training (PGY 1–5), actively engaged in neurosurgical rotations. 
The \textit{Student} group consisted of 14 students enrolled in a medical program, prior to starting residency. 
The study protocol received ethical approval from the McGill University Health Centre Research Ethics Board, Neurosciences-Psychiatry, and written informed consent was obtained from all participants.

Each participant was shown an annotated image indicating the specific subpial resection to be performed. 
The participants then completed three independent simulation trials. 
Each trial consisted of a subpial cortical resection, carried out as described in Section\,\ref{subsec:procedure}.
After each trial, participants completed a standardized questionnaire evaluating the platform's realism and educational value.
Results related to face and content validity derived from this questionnaire are reported in a separate study \cite{Almansouri2022} and are not included here.

\subsection{Platform setup}
\label{subsec:platform}


\begin{figure*}[t]
    \includegraphics[width=\textwidth]{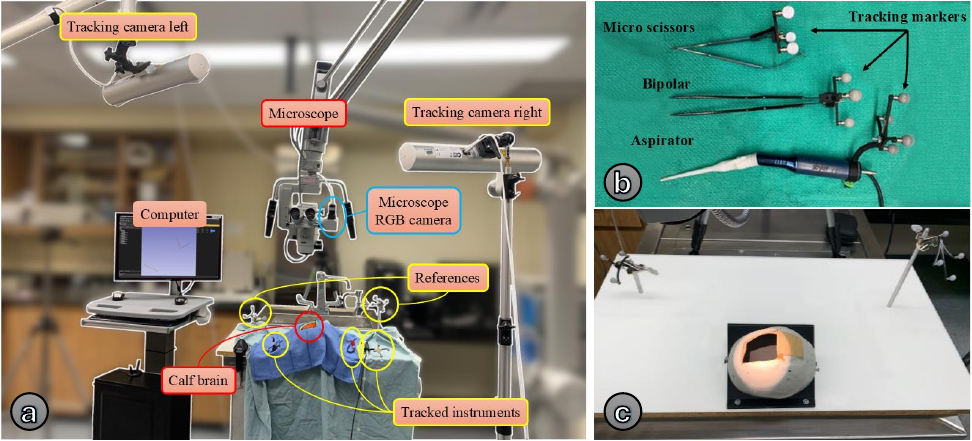}
    \caption{Overview of the simulation platform: (a) surgical (red), position tracking (yellow), and video recording (blue) components, (b) tracked instruments, and (c) setup showing the skull and craniotomy before installation of the calf brain.}
    \label{fig:setup}
\end{figure*}

Figure\,\ref{fig:setup}.a shows an overview of the simulation platform. 
The platform involves three main components: surgical field, position tracking, and video recording components.
The surgical components include the surgical microscope and instruments as well as an ex-vivo calf brain.
The calf brain was obtained from a local butcher and placed within a custom 3D-printed holder, ensuring that it remained stable at an ideal angle for the procedure (see Fig.\,\ref{fig:setup}.c). 
The holder was placed beneath a plastic cranial shell with a pre-cut window to simulate an off-midline craniotomy view. 
To enhance the realism of the setup, surgical towels and drapes were arranged to mimic the typical sterile field encountered in the operating room.
The complete setup was rigidly mounted on a flat wooden base to ensure stability during the procedure and enable repeatability between participants.

Instrument tracking in 3D space was achieved using a pair of infrared tracking cameras (FusionTrack 500, Atracsys LLC, Puidoux, Switzerland), with one camera oriented to capture the motion of instruments in the right hand and the other for the left hand.
During the procedure, three surgical instruments were tracked: the micro scissors, the bipolar forceps, and the ultrasonic aspirator (Fig.\,\ref{fig:setup}.b).
Instrument tracking was achieved by attaching lightweight carbon fiber rigid bodies (Atracsys LLC, Puidoux, Switzerland) to each tool.
We used custom-designed holders printed with polylactic acid (PLA) with 20\% infill to minimize weight and avoid disrupting the natural balance of the instruments.
This is done to preserve surgical fidelity and prevent adverse effects on bimanual performance.
Since the surgical task involved simultaneous use of both hands, two instruments were tracked concurrently--one in each hand. 
To ensure all tracking data were recorded within a unified coordinate reference frame, a dual reference system was used. 
A primary reference, oriented toward the right side, was designated as the global coordinate frame, while a secondary reference, facing the left, facilitated accurate instrument localization from both perspectives (see Fig.\,\ref{fig:setup}.c). 
The spatial transformation between the two references was established during a calibration phase to ensure consistent and accurate multi-instrument tracking throughout the simulation.

\subsection{Calibration}
\label{subsec:calibration}

The calibration process involved two key components: determining the position of the instrument tip (pivot calibration) and establishing the spatial transformation between the primary and secondary reference markers (reference calibration). 
Pivot calibration was performed by manually rotating each instrument around a fixed local point, allowing the tracking system to compute the three-dimensional coordinates of the instrument tip relative to the attached rigid body.
For the micro scissors and bipolar forceps, the instrument tip was defined as a single point located at the intersection of the tip of the jaws when fully closed, representing the effective point of contact during surgical manipulation.

To maintain a unified coordinate system on both sides of the surgical field, a calibration of the spatial relationship between the primary and secondary reference markers was also performed.
This was achieved by temporarily positioning a tracking camera directly above the reference markers and recording the transformation between them.
The resulting transformation matrix was saved and used throughout the procedure to align all tracking data within a common coordinate frame. 
This calibration method offers flexibility in setup, enabling the tracking cameras to be repositioned during simulation if necessary, without compromising the consistency or integrity of the recorded spatial data.

\subsection{Coordinate systems}
\label{subsec:transforms}

\begin{figure*}[t]
  \centering
  \includegraphics[width=0.9\linewidth]{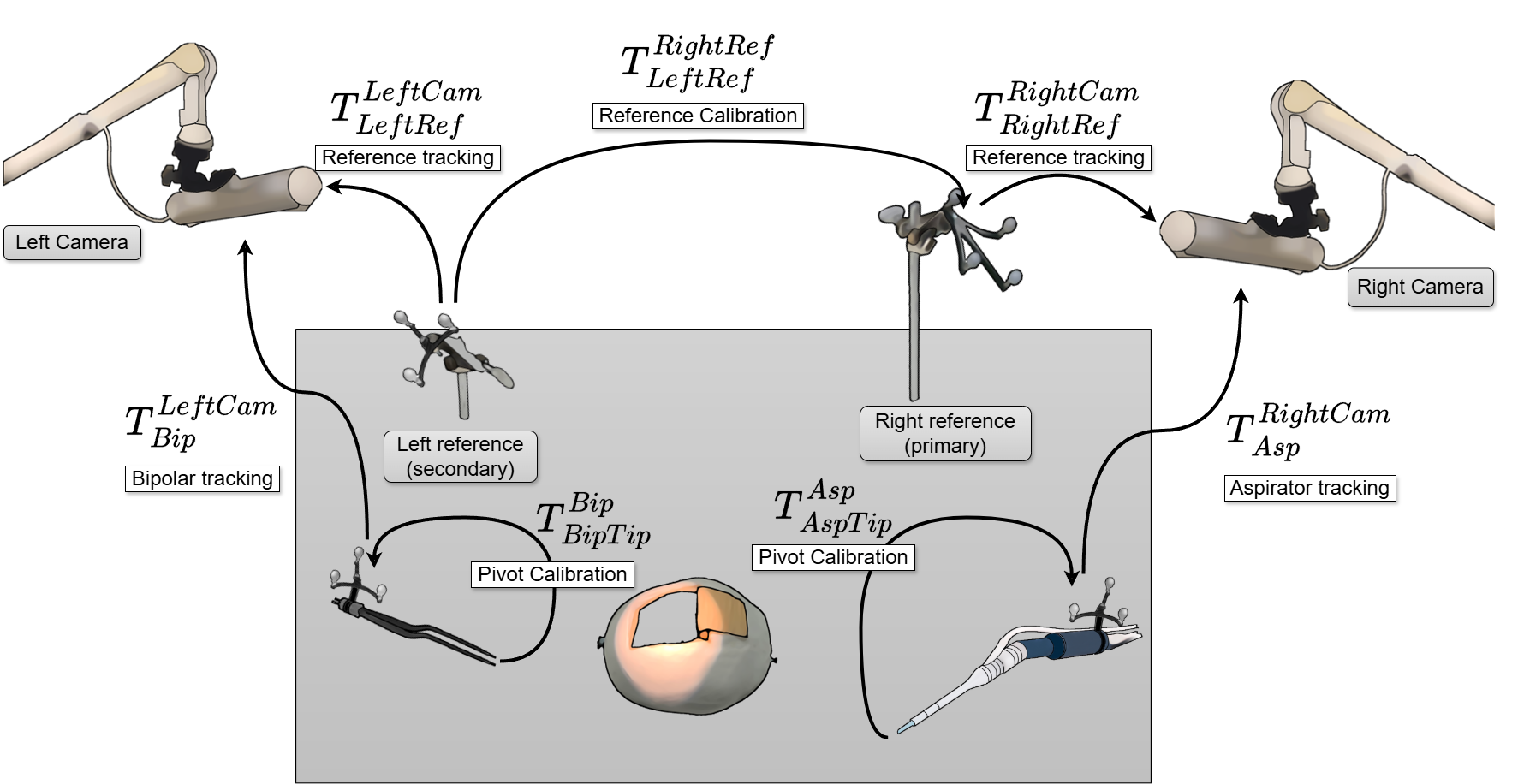}
  \caption{Coordinate system hierarchy and transform relationships in the simulation platform.}
  \label{fig:transforms}
\end{figure*}

The relationship between each transform used to track and align surgical instruments within a common coordinate system is illustrated in Fig.\,\ref{fig:transforms}.
The diagram outlines the hierarchical structure of transformations necessary to compute the position of each instrument tip (\textit{ToolTip}) in a global reference frame. 
The tracking system provides real-time estimates of the instrument's rigid body pose relative to the camera ($T_{\text{Tool}}^{\text{Cam}}$) and the reference marker pose relative to the camera ($T_{\text{Ref}}^{\text{Cam}}$).
The transformation between the instrument and its tip ($T_{\text{ToolTip}}^{\text{Tool}}$) is obtained by pivot calibration. 
Finally, the transform between the primary and secondary reference markers ($T_{\text{LeftRef}}^{\text{RightRef}}$), obtained from the reference calibration, enables the alignment of all tracked data within a unified coordinate system. 

\subsection{Data processing and annotation}
\label{subsec:annotations}

For data visualization and storage, we used 3D Slicer (https://www.slicer.org/) \cite{Fedorov2012}. 
A custom 3D Slicer extension was developed specifically for this study to automatically manage the transform hierarchy and streamline the simulation workflow.
Simulation trials were recorded in the form of Slicer-compatible sequence files, allowing frame-by-frame review and post-hoc analysis of surgical tool motion and positioning.

The PLUS Toolkit \cite{Lasso2014a} was used to interface with the tracking hardware, enabling real-time communication with both the FusionTrack 500 and Polaris camera systems. 
The toolkit facilitates synchronized acquisition of six-degree-of-freedom (6-DOF) pose data from multiple tracked instruments and reference markers, ensuring consistency in spatial and temporal coherence across all data streams. 
Tracking information was transmitted to 3D Slicer via the OpenIGTLink protocol \cite{Tokuda2009}, an open-source network communication standard specifically developed for image-guided therapy applications. 

In addition, microscope video footage of each simulation session was recorded using the native software provided by FLIR Systems, which handled the acquisition of high-resolution videos at an average of 80 frames per second. 
These videos served as a visual reference to complement the tracked data and enabled synchronized assessment of instrument motion, user interaction, and surgical behavior during each trial.
Since video recordings and tracking data were captured using separate software systems, they did not share a common timestamp. 
To address this, a manual post-processing synchronization step was performed for each trial. 
At the beginning of the procedure, participants were instructed to bring the tips of the bipolar forceps and the aspirator into contact. 
This specific gesture was used as a temporal marker to align the video and tracking timelines.
In addition to temporal synchronization, each video was manually annotated to identify the time segments during which each instrument was actively in use. 
These annotated intervals served as a reference for subsequent time-based analyses.

\subsection{Metrics}
\label{subsec:metrics}

To evaluate instrument kinematics, we extracted three main categories of metrics summarized in Table\,\ref{tab:metrics}: motion-based, time-based, and efficiency/coordination metrics. 

\begin{table*}[t]
\caption{Summary of the metrics extracted from instrument tracking and video recording.}
\centering
\begin{tabular}{cp{0.225\textwidth} p{0.7\textwidth}}
\toprule
 & \textbf{Metrics} & \textbf{Description} \\
\midrule
\multicolumn{2}{l}{\textbf{Motion-based Metrics}} & \\
& Average Velocity & The mean speed at which each instrument tip moves throughout the entire trial, measured in (mm/s). \\
& Average Acceleration & The mean rate of change in \textit{velocity} of the instrument tip during the procedure, measured in (mm/s²). \\
& Average Jerk & The mean rate of change of \textit{acceleration}, used to quantify the smoothness of instrument motion, in (mm/s³). \\
& Normalized Path Length &  The cumulative distance traveled by the instrument tip normalized over the total time of the instrument usage ($T_\textit{Usage}$), measured in (mm/s). \\

\midrule
\multicolumn{2}{l}{\textbf{Time-based Metrics}} & \\
& Usage Time ($T_\textit{Usage}$) & The total duration during which the instrument was manually annotated as being in active use, regardless of tracking succes, measured in (s). \\
& Tracking Time ($T_\textit{Track}$) & The amount of time the instrument was successfully detected and tracked by the system, measured in (s). \\
& Captured Time ($T_\textit{Capt}$) & The overlap in time during which the instrument was both annotated as in use and successfully tracked, representing reliable tracking captured during active manipulation, measured in (s). \\
& Percentage of Captured Time & The proportion of in-use time that was effectively tracked, used to evaluate system tracking efficiency, in (\%). \\

\midrule
\multicolumn{2}{l}{\textbf{Efficiency \& Coordination}} & \\
& Average Separation Distance (ASD) & The mean spatial distance between the tips of the instruments used in the dominant and non-dominant hands during bimanual tasks, measured in (mm). \\
& Efficiency Index (EI) & The ratio of the time the ultrasonic aspirator was used during the resection to the total procedure time, independent of tracking data. $\text{EI}\in [0,1]$ \\
& Coordination Index (CoordIdx) & The proportion of time the bipolar and aspirator were used together relative to the time the aspirator was used alone, assessing coordinated bimanual performance. $\text{CoordIdx}\in [0,1]$ \\
& Bimanual Usage Time ($\textit{BiT}_\textit{Usage}$) & Measure how long both instruments were in use during bimanual activity, measured in (s). \\
& Bimanual Tracked Time ($\textit{BiT}_\textit{Track}$) & Measure how long both instruments were tracked during bimanual activity, measured in (s). \\
& Bimanual Captured Time ($\textit{BiT}_\textit{Capt}$) & Measure how long both instruments were simultaneously in use and tracked during bimanual activity, in (s). \\

\bottomrule
\end{tabular}
\label{tab:metrics}
\end{table*}

Distance-based metrics are derived directly from the positional data of the instrument tip. 
These include:
\begin{itemize}
    \item The first, second and third position derivatives--\textit{velocity}, \textit{acceleration}, and \textit{jerk}, respectively--which characterize the dynamics of instrument movement. 
    \item \textit{Normalized Path Length}: defined as the cumulative distance traveled by the instrument tip during a trial. Since longer procedures naturally lead to longer path lengths, the path length was normalized by the total time the instrument was in use, to enable for fair comparisons across trials of varying durations. 
\end{itemize}

Time-based metrics quantify how effectively the tracking system captures periods of instrument use.
These metrics account for four possible scenarios during a trial: (1) the instrument is in use but not tracked, (2) the instrument is tracked but not actively used (e.g., resting on the table), (3) the instrument is both tracked and in use, and (4) the instrument is neither tracked nor in use. 
To characterize these scenarios, we defined three temporal measures:
\begin{itemize}
    \item \textit{Usage Time ($T_\textit{Usage}$)}: Obtained through manual annotation of video recordings, $T_\textit{Usage}$ indicates the ground-truth period during which the instrument was actively used, regardless of tracking performance.  
    \item \textit{Tracking Time ($T_\textit{Track}$)}: Computed from the output of the tracking system, $T_\textit{Track}$ reflects the duration for which the instrument was successfully tracked.  
    \item \textit{Captured Time ($T_\textit{Capt}$)}: Defined as the intersection of $T_\textit{Usage}$ and $T_\textit{Track}$, $T_\textit{Capt}$ captures the duration during which the instrument successfully tracked while being actively used. This metric serves as an indicator of the effectiveness of the tracking system during active manipulation.
\end{itemize}

Finally, efficiency/coordination metrics were developed to evaluate higher-order psychomotor and cognitive skills involved in bimanual neurosurgical tasks. 
These metrics are particularly relevant for assessing the planning, synchronization, and execution required during complex procedures such as subpial corticectomy. 
These metrics include:
\begin{itemize}
    \item \textit{Average Separation Distance (ASD)}: represents the mean distance between the tips of the instruments held in each hand. 
    This metric has been shown to correlate with bimanual coordination \cite{Yilmaz2022}, with more consistent and controlled separation, often reflecting a higher level of skill in maintaining spatial awareness and tool interaction.
    \item \textit{Efficiency Index (EI)}: quantifies the proportion of time the aspirator was actively used during cortical resection relative to the total duration of the task. 
    This metric is intended to capture the interaction between cognitive planning and motor execution, providing information on the trainee's ability to optimize the sequence and timing of surgical actions.
    \item \textit{Coordination Index (CoordIdx)}: reflects the quality of bimanual interaction by calculating the ratio of time the aspirator and bipolar forceps were used simultaneously to the time the aspirator was used alone. 
    A higher CoordIdx indicates more effective support of the dominant-hand instrument by the non-dominant hand, an indicator of advanced surgical dexterity and coordinated workflow.
    \item In parallel to the single-instrument time-based metrics, we also defined \textit{Bimanual Usage Time ($\textit{BiT}_\textit{Usage}$)}, \textit{Bimanual Tracking Time ($\textit{BiT}_\textit{Track}$)}, and \textit{Bimanual Captured Time ($\textit{BiT}_\textit{Capt}$)}, which respectively denote the durations during which both instruments were manually annotated as in use, successfully tracked, and simultaneously tracked while in use. 
    These measures offer additional insight into the trainee’s ability to maintain coordinated and sustained bimanual engagement throughout the procedure.
\end{itemize}

\subsection{Statistical analysis}
\label{subsec:stats}


Statistical analyses were conducted in R v4.5.1\,\cite{R2021}.
For each instrument, average performance metrics were compared across four expertise levels (Student, Junior, Senior, Expert) using one-way analysis of variance (ANOVA) mixed-effects models, with level of expertise included as a fixed factor and participant as a random factor to account for repeated measures within participants across three trials. 
Assumptions of ANOVA model errors, including Normality and homogeneity of variance, as well as the presence of outliers or influential observations were assessed by graphical examination of model residuals. 
When warranted, post-hoc pairwise comparisons of mean differences were performed, and p-values were adjusted for multiple testing using the Tukey method. 
Results are reported as estimated mean differences and 95\% confidence interval (CI). 
All statistical tests of hypothesis were two-sided and performed at the 5\% level of significance. 

\section{Results}
\label{sec:results}

A total of 47 participants were enrolled in the study, each completing three simulation trials, resulting in a total of 141 trials. 
Five trials were excluded from the analysis due to data loss or corruption caused by recording errors.
This involved two first-trial, one second-trial, and two third-trial sessions from five different participants.
The remaining 136 trials were retained for subsequent analysis.
Detailed statistical results are provided in the supplementary material (Suppl1).

\subsection{Tracking performance}
\label{subsec:tracking_results}


Table\,\ref{tab:tracking_perf} summarizes the overall tracking performance of the simulation platform per instrument. 
An average of 81\% ± 25\% of the instrument usage time was successfully captured across all instruments and participants, indicating generally reliable tracking during periods of active instrument use.
Tracking losses mainly occurred due to instrument marker occlusions from the cameras' view, caused by non-standard or large variation of instrument handling during the procedure and/or physical obstructions by participants.

\begin{table}[t]
\centering
\caption{Summary of mean (standard deviation) for tracking performance metrics per instrument}
\label{tab:tracking_perf}
\begin{tabular}{clccc}
\toprule
\multicolumn{2}{l}{\textbf{Tracking data}} & \textbf{Bipolar} & \textbf{Aspirator} & \textbf{Scissors} \\
\midrule
\multicolumn{2}{l}{Success rate (\%)} & 87.96 (19.84) & 73.79 (22.25) & 79.40 (30.34) \\
\multicolumn{2}{l}{Time (s)} &&&\\
& Captured  & 353.2 (212.5) & 220.8 (148.8) & 85.9 (84.4)\\
& Annotated & 404.8 (220.2) & 312.0 (207.4) & 100.7 (91.3)\\
\bottomrule
\end{tabular}
\end{table}

\subsection{Motion-based metrics}
\label{subsec:motion_results}

\begin{figure*}[ht]
    \centering
    \includegraphics[width=0.75\linewidth]{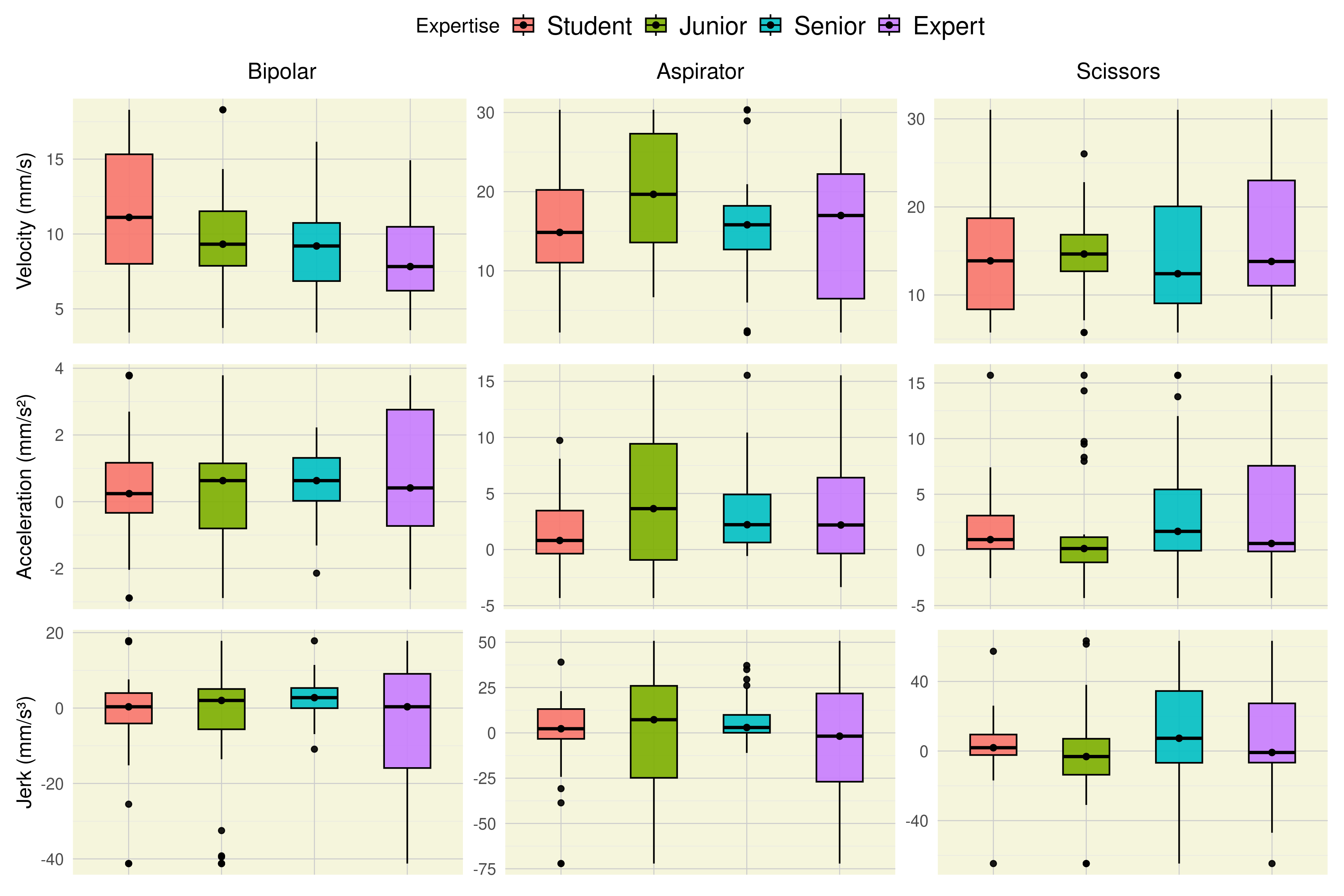}
    \caption{Motion-based metric results for \textit{velocity}, \textit{acceleration} and \textit{jerk}.}
    \label{fig:vaj_results}
\end{figure*}

Figure\,\ref{fig:vaj_results} shows results for the average \textit{velocity}, \textit{acceleration}, and \textit{jerk} calculated for each instrument type: bipolar forceps, ultrasonic aspirator, and micro scissors for all simulation trials. 
These metrics provide a global measure of the instrument's motion dynamics over the entire duration of the trial session.
No statistically significant differences were found for any of the global motion-based metrics regardless of the instrument.


\begin{figure*}
    \centering
    \includegraphics[width=0.75\linewidth]{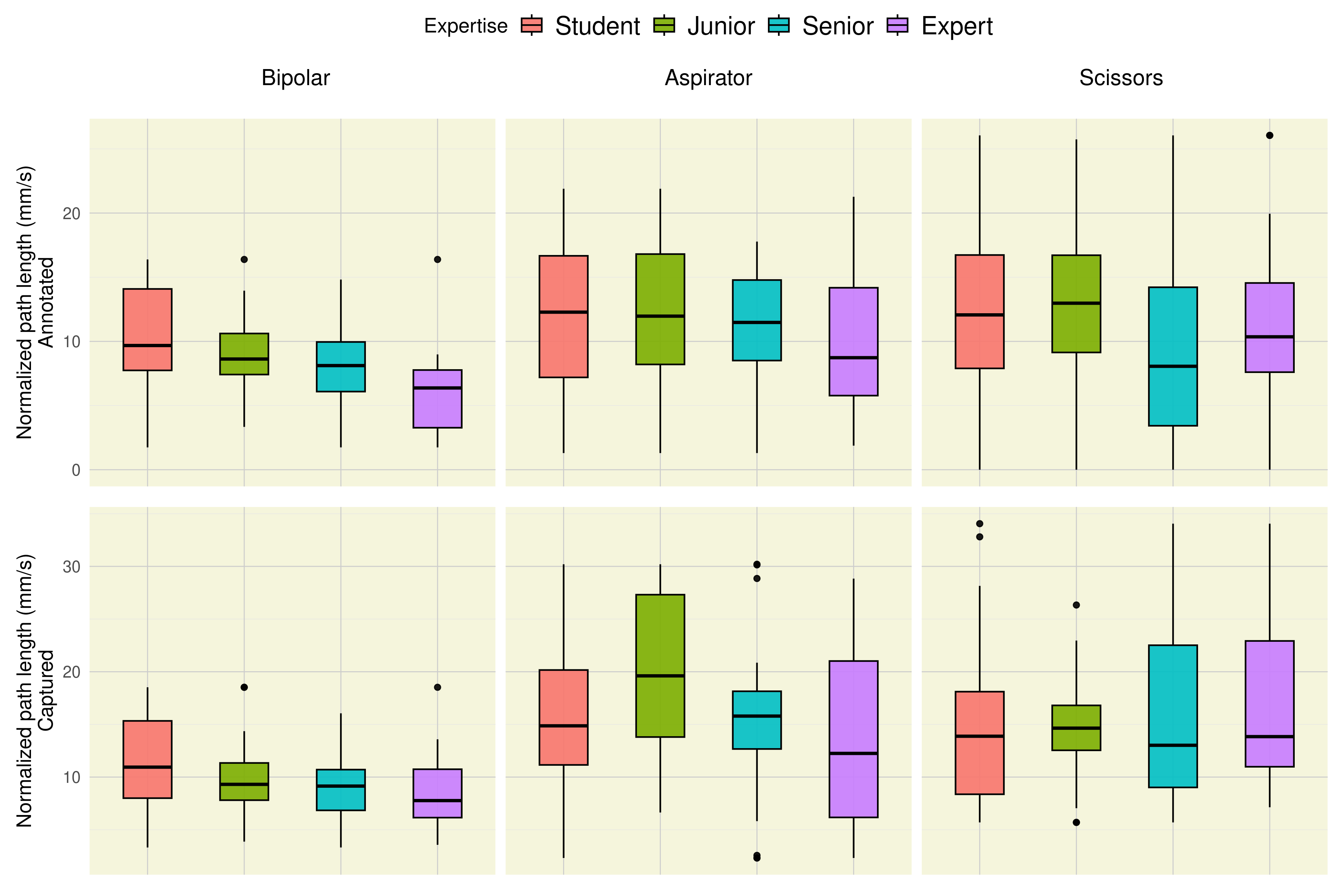}
    \caption{The Normalized Path Length denotes the distance traveled by the instrument divided by its usage time: Annotated normalized path length was measured using Usage Time ($T_\textit{Usage}$), and Captured normalized path length was measured using Captured Time ($T_\textit{Capt}$).}
    \label{fig:npl}
\end{figure*}

Results for the normalized path length are shown in Fig.\,\ref{fig:npl}. 
Annotated normalized path length values were obtained using manually annotated video segments during active instrument use; these serve as reference standards for comparison with tracking performance.
On the other hand, the captured normalized path length values were calculated during segments where the instruments were used and tracked simultaneously, reflecting the system's ability to accurately capture this metric.
The latter was used in the statistical analysis.
The median (interquartile range [IQR: 25$^\text{th}$percentile--75$^\text{th}$percentile]) ratio of captured-to-annotated normalized path length ($nPL_{\text{captured}} / nPL_{\text{annotated}}$) is $1.06$\,[1.0--1.13], $1.26$\,[1.11--1.58] and $1.05$\,[1.0--1.29], for bipolar, aspirator and scissors, respectively.
There is a near 1-to-1 correspondence between manual annotation and tracking-based normalized path length measurements, with a slight overestimation for aspirator normalized path length.
Statistical analysis of the normalized path length did not show any significant differences across expertise levels or instruments.

\subsection{Time-based metrics}
\label{subsec:time_results}

\begin{figure*}[t]
    \centering
    \includegraphics[width=0.75\linewidth]{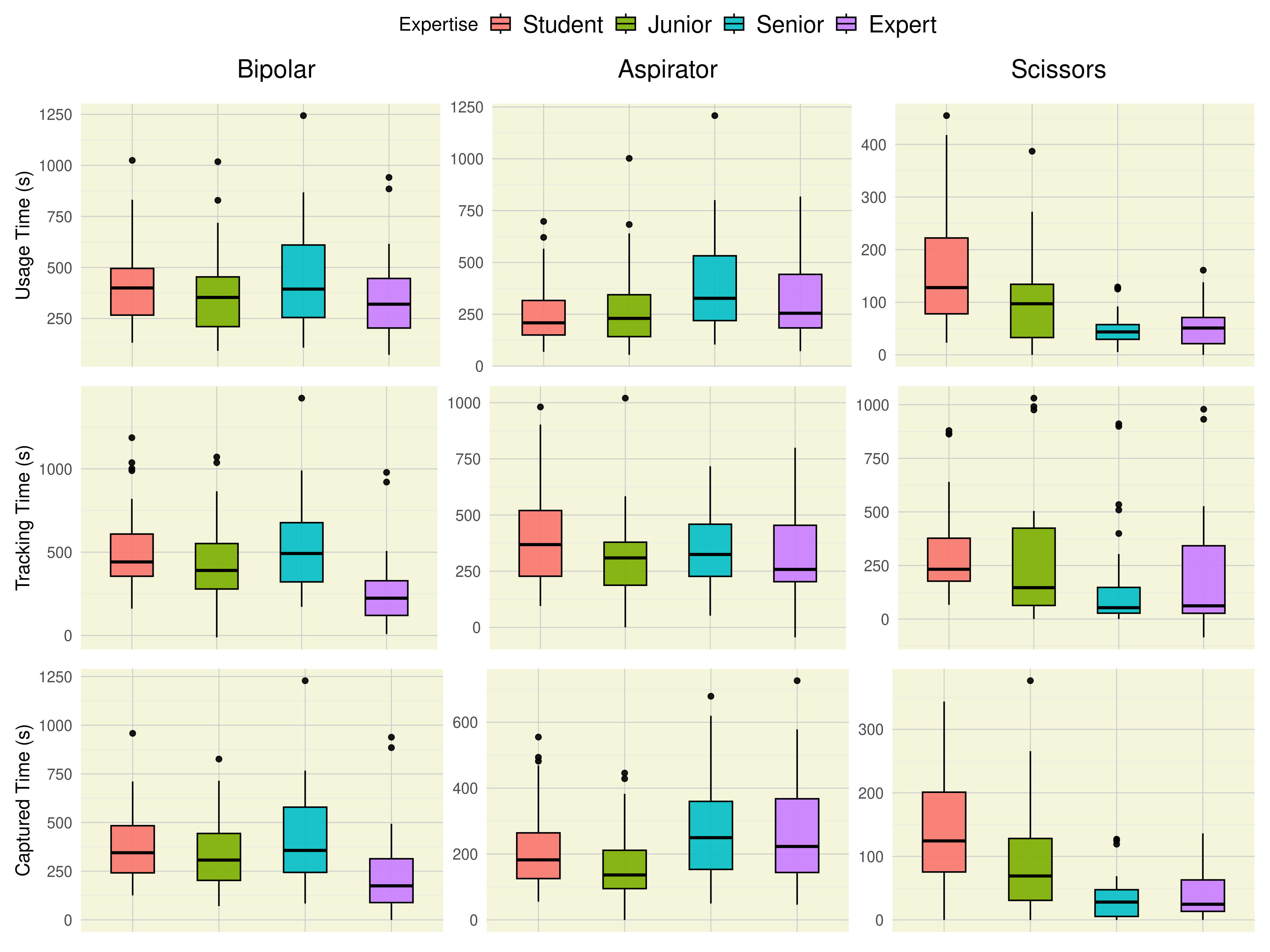}
    \caption{Single instrument time results: Usage Time ($T_\textit{Usage}$) is derived from manual video annotations, Tracking Time ($T_\textit{Track}$) denotes successful system tracking, and Captured Time ($T_\textit{Capt}$) denotes effective tracking within the manually annotated window.}
    \label{fig:time}
\end{figure*}

Results of time-based metrics are shown in Fig.\,\ref{fig:time}. 
These metrics capture the period of time during which instruments were actively used (manually annotated from videos), successfully tracked, and simultaneously used and tracked (i.e., captured time), providing insight into both system performance and participant behavior across different levels of expertise.
The median [IQR: 25$^\text{th}$percentile--75$^\text{th}$percentile] ratio of captured-to-usage time ($T_\textit{Capt}/T_\textit{Usage}$) is $0.95$\,[1.0--0.88], $0.78$\,[0.9--0.61] and $0.94$\,[1.0--0.75] for bipolar, aspirator, and scissors, respectively.

Statistical analysis revealed differences in usage time for the scissors and aspirator instruments.
Regarding scissors time, Students used scissors significantly more than both Experts and Seniors, with mean differences of about 101.46\,s [CI: 35.65, 167.28] ($p = 0.001$) and 106.34\,s [CI: 46.77, 165.91] ($p < 0.001$), respectively. 
Although Experts tended to use scissors less than Juniors (–54\,s, $p = 0.15$) and Juniors tended to be used scissors less than Students (–48\,s, $p = 0.11$), these effects did not reach statistical significance after Tukey adjustment. 
The Junior-Senior contrast did not show a significant difference as well ($p = 0.057$).
Regarding aspirator usage time, the Tukey-adjusted comparisons showed that Juniors used the aspirator significantly longer than Seniors (mean difference = 125.8\,s [CI: 4.68, 246.8], $p = 0.039$). 
No other pairwise contrasts reached significance.


\begin{figure}[t]
    \centering
    \includegraphics[width=0.5\linewidth]{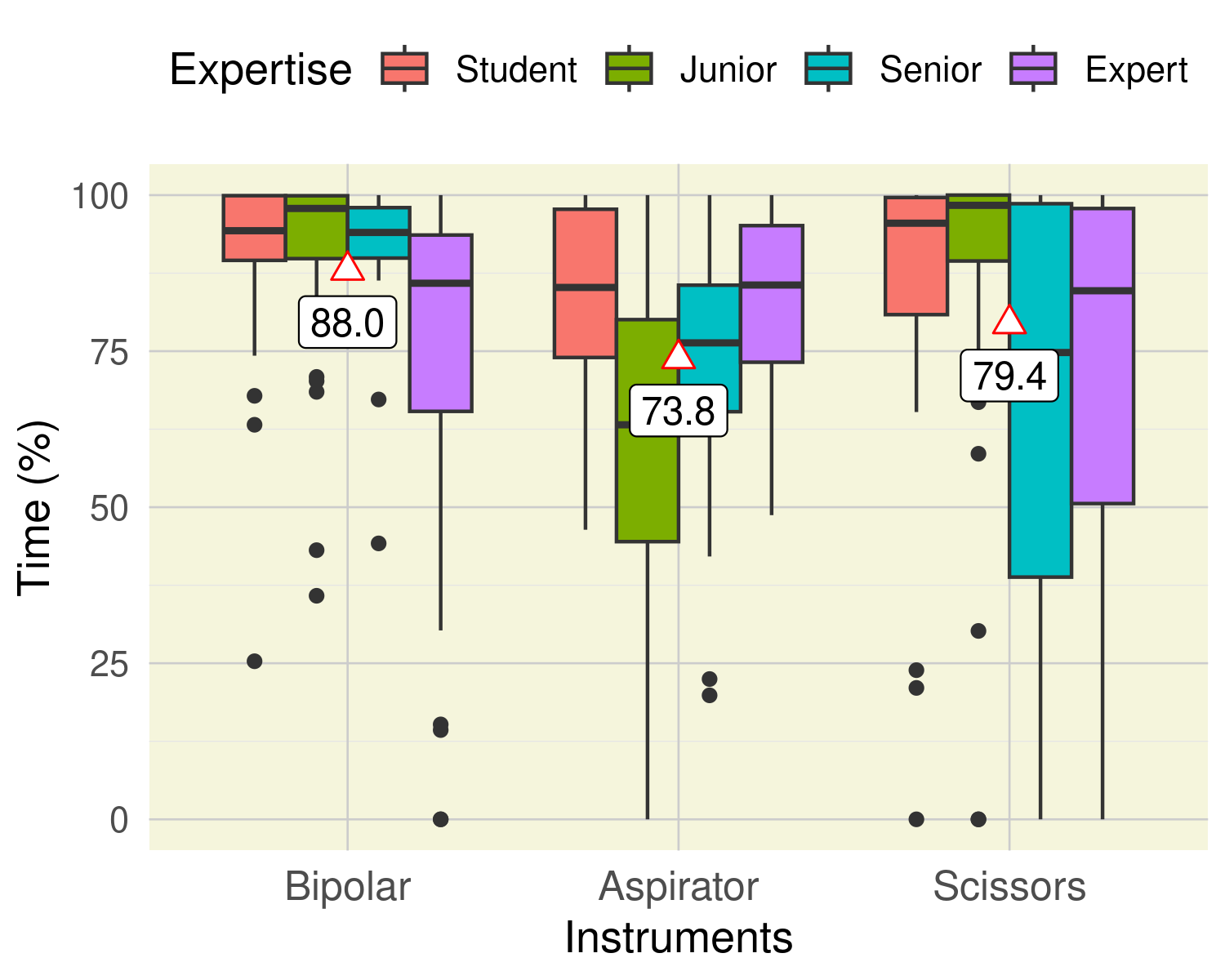}
    \caption{Captured-to-usage time ratios: white triangles indicate average \% per instrument across all expertise levels.}
    \label{fig:time_percentage}
\end{figure}

Figure\,\ref{fig:time_percentage} illustrates the results of ratios between the captured time and the usage time.
This provides insight into the system’s ability to reliably track instruments during periods of active use.
A higher ratio indicates greater tracking fidelity within the time segments where the instrument is being manipulated by the participant. 

\subsection{efficiency/coordination}
\label{subsec:efficiency_results}

\begin{figure}[t]
    \centering
    \includegraphics[width=.6\linewidth]{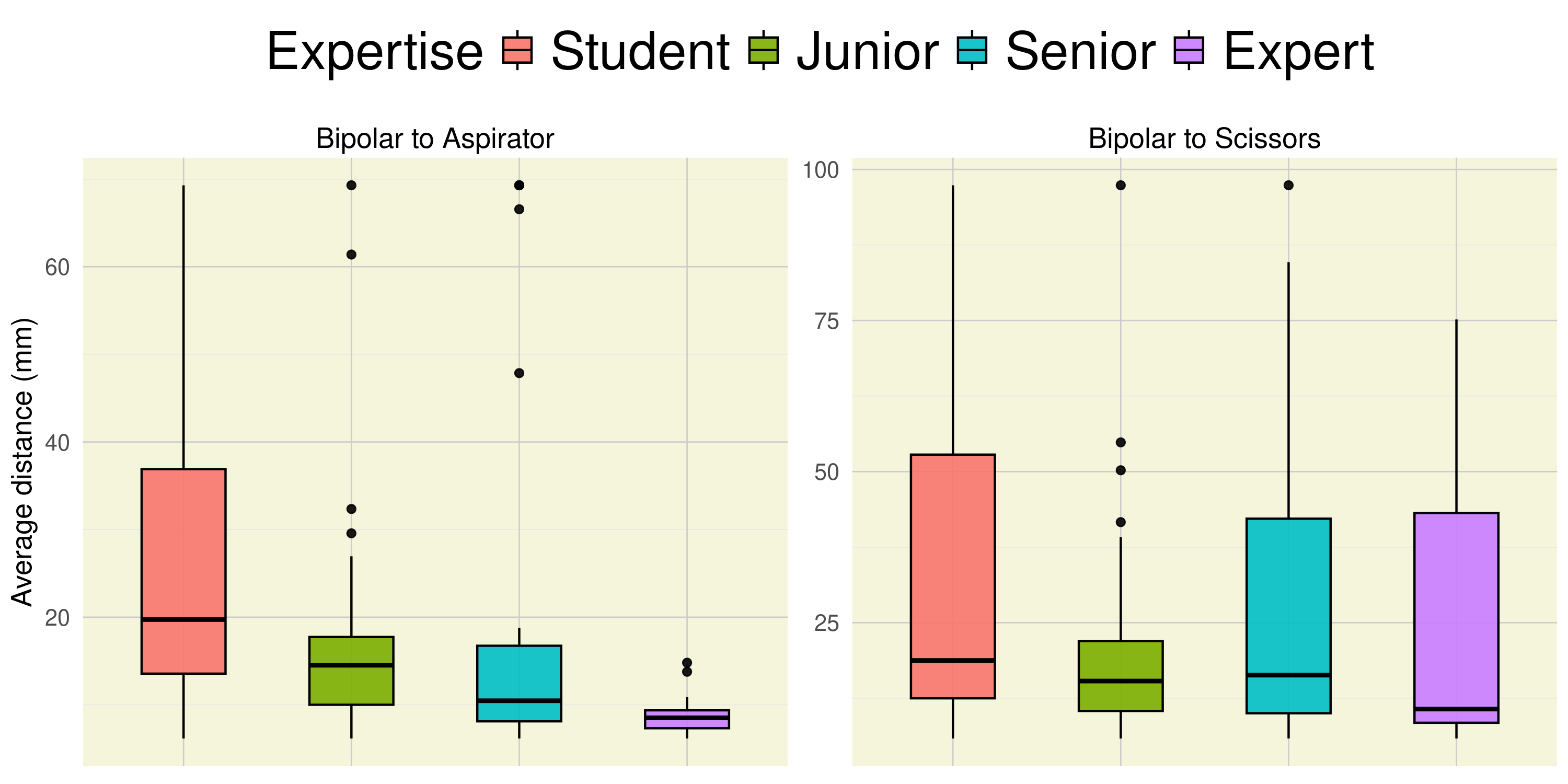}
    \caption{Results of the Average Separation Distance (ASD) between instrument tips in dominant and non-dominant hands.}
    \label{fig:bi_distance}
\end{figure}

Figure\,\ref{fig:bi_distance} shows the average distance between the tips of two instruments.
This metric is computed during intervals in which both instruments were simultaneously used and successfully tracked, thereby eliminating periods of instrument inactivity. 
The metric serves as a proxy for the quality of spatial coordination during bimanual manipulation.  
A statistically significant difference was found in the distance between the bipolar forceps and the aspirator. 
Post hoc Tukey-adjusted comparisons revealed that Students maintained a significantly greater distance than both Experts (mean difference of 20.6\,mm [CI: 5.7, 35.5], $p = 0.003$) and Seniors (mean difference of 15.3\,mm [CI: 2.1, 28.4], $p = 0.018$). 
No significant differences were observed among Experts, Juniors, and Seniors.

\begin{figure}[t]
  \centering
  \begin{subfigure}{0.45\textwidth}
    \centering
    \includegraphics[width=0.85\textwidth]{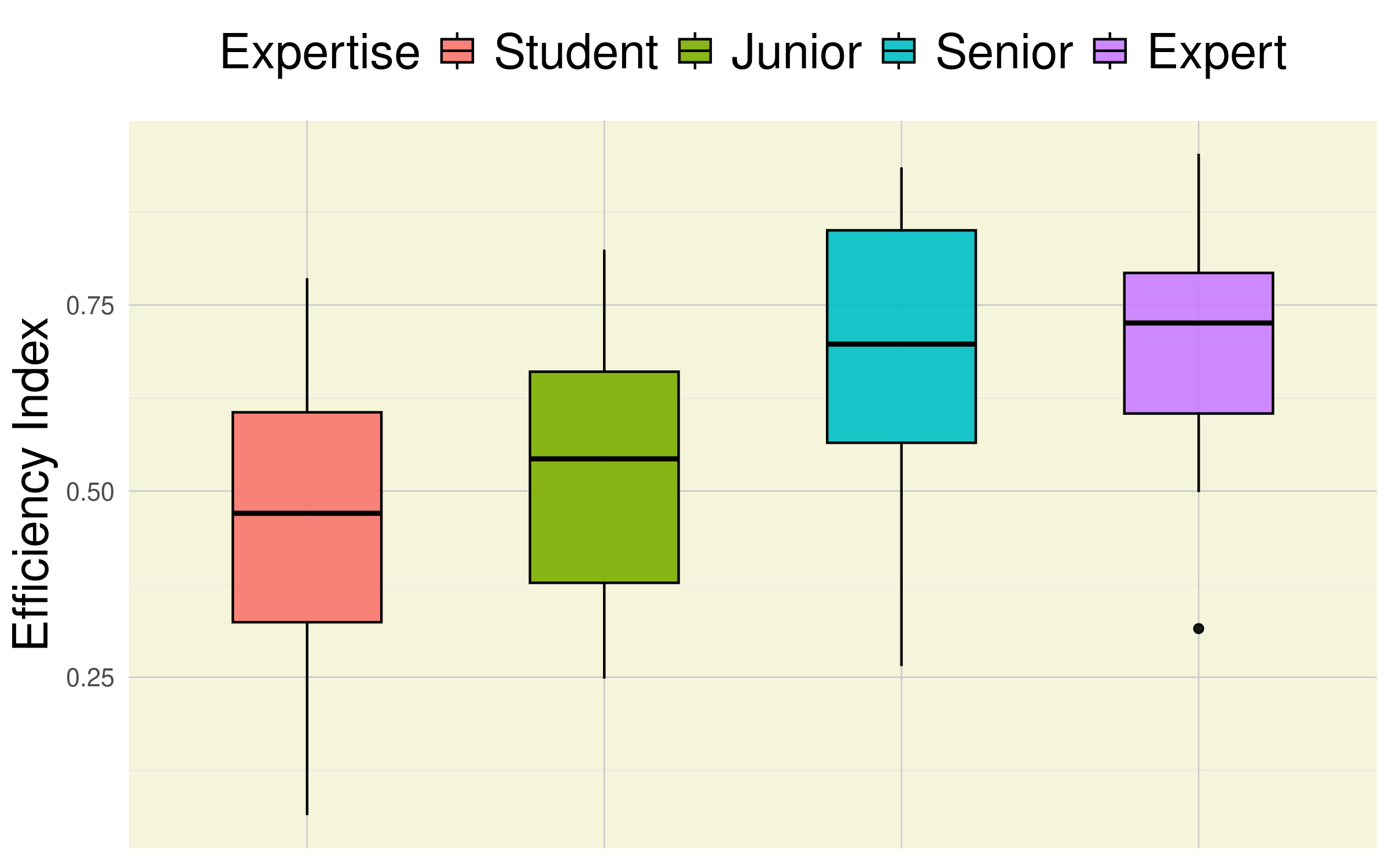}
  \end{subfigure}%
  
  \begin{subfigure}{0.45\textwidth}
    \centering
    \includegraphics[width=0.9\textwidth]{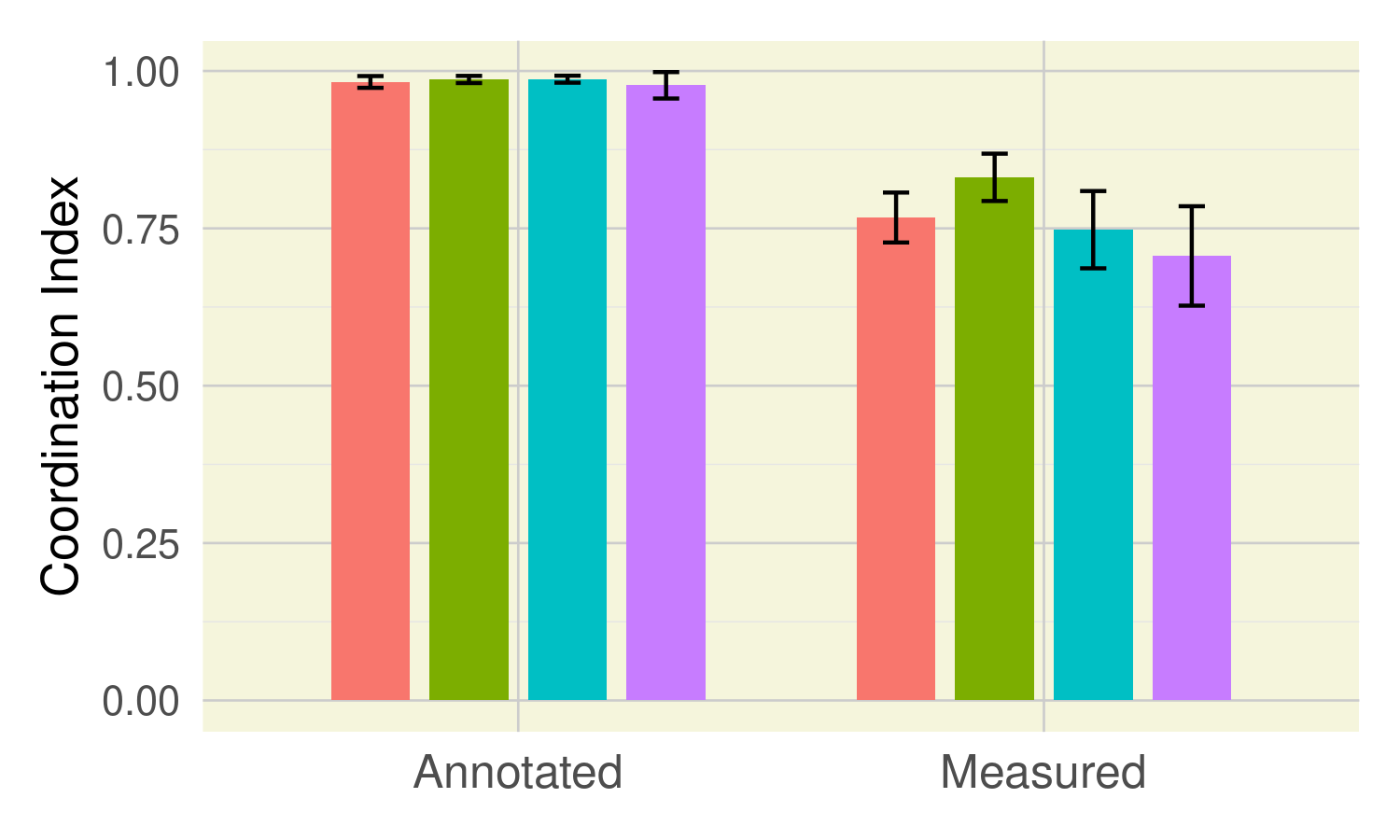}
  \end{subfigure}
  \caption{Results of efficiency/coordination: (top) Efficiency Index is the ratio of time aspirator was use during resection to total procedure time, (bottom) Coordination Index is the ratio of time both bipolar and aspirator were used together to the time the aspirator was used alone.}
  \label{fig:ei_ci}\hfill
\end{figure}

Figure\,\ref{fig:ei_ci} (top) presents the results for the Efficiency Index (EI), which quantifies the proportion of time the aspirator was actively used during resection relative to the total duration of the task.
Statistical analysis reveals that Experts demonstrated a higher EI compared to both Students (mean difference of 0.24 [CI: 0.08, 0.41], $p = 0.002$) and Juniors (mean difference of 0.18 [CI: 0.01, 0.35], $p = 0.029$).
Seniors also showed a significantly higher EI than Students (mean difference of 0.20 [CI: 0.06, 0.36], $p = 0.004$).
No significant differences were observed between Experts and Seniors ($p = 0.95$), Juniors and Seniors ($p = 0.066$), or between Juniors and Students ($p = 0.63$). 

In Fig.\,\ref{fig:ei_ci} (bottom), results of the Coordination Index (CoordIdx) are shown.
The CoordIdx represents the proportion of aspirator use time that overlaps with the simultaneous use of the bipolar forceps.
No statistically significant differences were found in CoordIdx between the expertise groups.
In addition, we compare the CoordIdx values calculated using manually annotated usage times ($T_\textit{Usage}$), serving as a reference, with those calculated using time captured by the tracking system ($T_\textit{Capt}$). 
The mean (standard deviation) score for annotated CoordIdx was $0.98 (0.06)$, while the average score for captured CoordIdx was $0.77 (0.29)$.

\begin{figure*}[t]
    \centering
    \includegraphics[width=.5\linewidth]{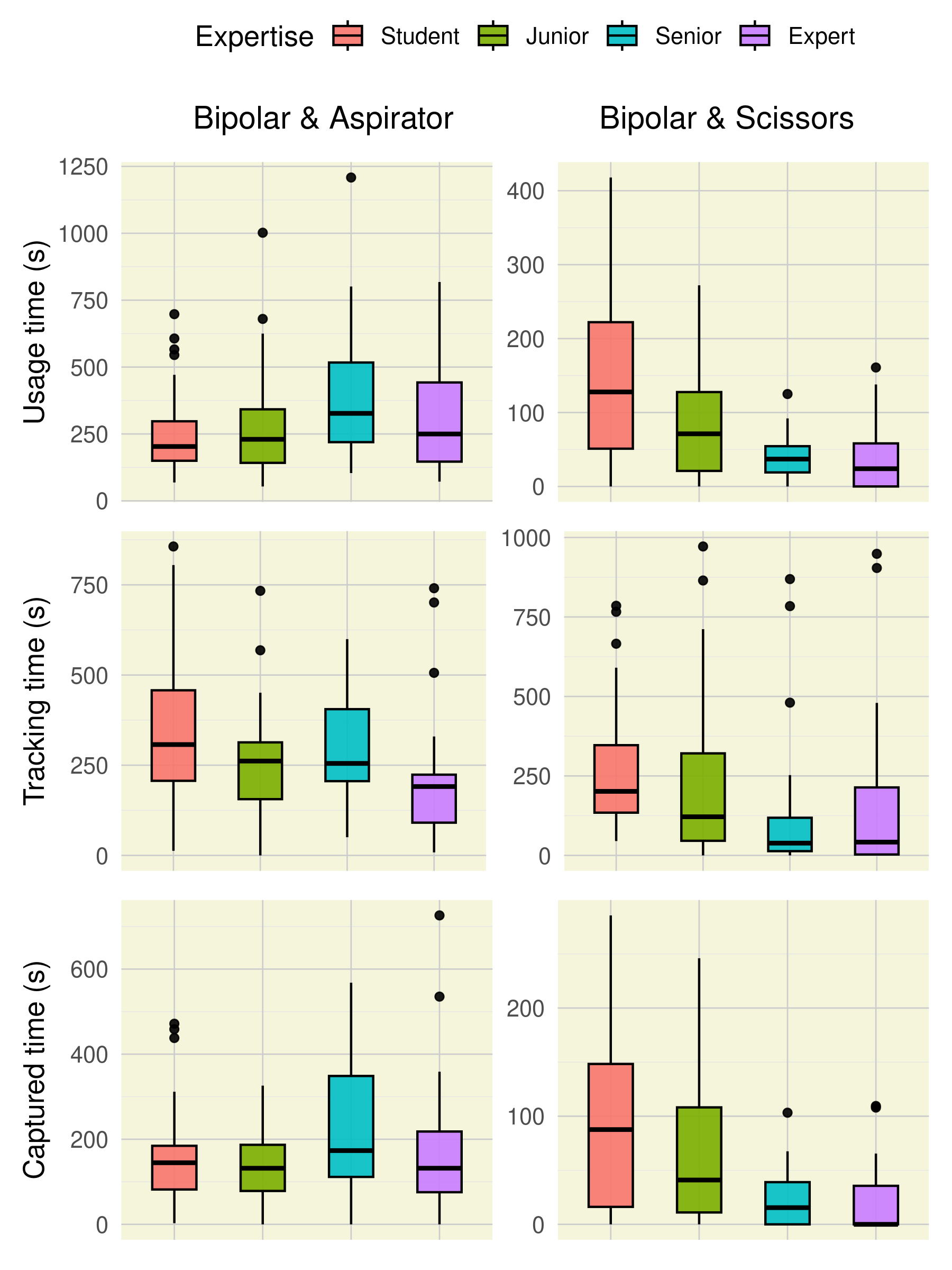}
    \caption{Bimanual time results consider simultaneous instrument usage: Usage Time ($\textit{BiT}_\textit{Usage}$) via manual annotations, Tracking Time ($\textit{BiT}_\textit{Track}$) via system tracking, and Captured Time ($\textit{BiT}_\textit{Capt}$) via system tracking during annotation periods.}
    \label{fig:bi_time}
\end{figure*}

Results of the time-based bimanual metrics are presented in Fig.\,\ref{fig:bi_time}.
The simultaneous usage time of bipolar forceps and scissors revealed significant differences between levels of expertise. 
Students used both instruments simultaneously for significantly longer periods compared to Experts (mean difference of 77.3\,s [CI: 15.5, 139.0], $p = 0.009$) and Seniors (mean difference of 75.4\,s [CI: 19.4, 131.3], $p = 0.004$).
No significant differences were observed between Experts and Juniors ($p = 0.21$), Experts and Seniors ($p \approx 1.0$), Juniors and Seniors ($p = 0.17$), or Juniors and Students ($p = 0.39$).

\section{Discussion}
\label{sec:discussion}

This study demonstrates the feasibility of tracking surgical instruments during subpial resection, with the system successfully capturing instrument motion in 81\% of the usage time.

\subsection{Efficiency of metrics in differentiating expertise levels}

\begin{table}[t]
    \centering
    \caption{Summary of significant differences for metric-instrument results with the associated expertise-level pairs.}
    \label{tab:significant_metrics}
    \begin{tabular}{llccccr}
        \toprule
        \textbf{Metric} & \textbf{Instrument} & \multicolumn{4}{l}{\textbf{Expertise-level pair}} & \multicolumn{1}{l}{\textbf{p-value}} \\ \bottomrule
        Time & Aspirator & & Jr & Sr & & 0.04 \\ \cline{2-7}
         & Scissors & S &  & Sr & & $<0.01$ \\ \cline{3-7}
         & & S & & & Exp & $<0.01$ \\ \hline
        Efficiency Index & - & S & & Sr & & $<0.01$ \\ \cline{3-7}
         & & S & & & Exp & 0.02 \\ \cline{3-7}
         & & & Jr & & Exp & 0.03 \\ \hline
        ASD & Bipolar-Aspirator & S & & Sr & & $0.02$ \\ \cline{3-7}
         & & S & & & Exp & $<0.01$ \\ \hline
        Bimanual Time & Bipolar-Scissors & S & & Sr & & $0.02$ \\ \cline{3-7}
         & & S & & & Exp & $<0.01$ \\ \bottomrule
         \multicolumn{7}{l}{Expertise-level -- S: Student, Jr: Junior, Sr: Senior, Exp: Expert.}\\
         \multicolumn{7}{l}{ASD: Average Separation Distance.}\\
    \end{tabular}
\end{table}

We analyzed instrument trajectories to derive motion-based, time-based, and bimanual efficiency metrics to differentiate participants by surgical expertise. 
Table\,\ref{tab:significant_metrics} summarizes the key metrics across different expertise levels that demonstrated statistically significant differences.

Within the set of metrics examined, none of the motion-based metrics demonstrated discriminative capabilities to distinguish between different levels of surgical expertise, regardless of the instrument used. 
In contrast, time-based metrics demonstrated discriminative capability for two instruments. 
Specifically, aspirator usage time tended to increase with expertise, with a significant difference observed between the Junior and Senior groups. 
Conversely, scissors usage time decreased significantly as expertise level increased.
It is worth noting that scissors usage may be influenced by strategic decisions in task execution, thereby reflecting underlying differences in procedural planning and task familiarity across training levels. 
However, the limited overall use of scissors in this task warrants cautious interpretation of these results.
Future studies should aim to validate these observations in more complex procedures or in settings involving prolonged instrument engagement.

We also investigated a set of custom-designed bimanual metrics, encompassing both motion-based and time-based features that quantify the coordinated use of two instruments simultaneously. 
Bipolar–scissors simultaneous usage time differentiated between lower- and higher-level groups, distinguishing Students from both Seniors and Experts.
Participants with greater surgical expertise tended to use both instruments simultaneously for longer durations, which may reflect improved bimanual coordination and task integration.
This trend is further supported by the Efficiency Index, which captures the proportion of time during which the aspirator was actively used. 
Interestingly, although the Efficiency Index is calculated solely based on aspirator usage, it exhibits a similar discriminative pattern to the simultaneous usage time of the bipolar–scissors pair.
This trend is also observed in the average separation distance between the tips of the bipolar and aspirator instruments, with more experienced participants maintaining a shorter and more consistent distance during bimanual manipulation. 
A reduced separation distance may indicate more deliberate and coordinated control of both instruments, reinforcing the notion that bimanual metrics can provide meaningful insights into surgical expertise.

We also note that no metric successfully discriminated between students and junior residents, nor between senior residents and neurosurgeons. 
The lack of separation between students and juniors may stem from the high variability within the junior group, which encompasses residents from post-graduate years 1 through 5. 
At the upper end of training, the apparent plateau between seniors and neurosurgeons may reflect the convergence of psychomotor performance at advanced stages, where more subtle cognitive or procedural factors, not captured by instrument motion characteristics, become more decisive in distinguishing expertise.

\subsection{Study limitations}

This study presents several limitations that should be considered when interpreting the findings. 
Tracking the ultrasonic aspirator was the most challenging.
This is largely attributable to the way the aspirator was used, often beneath the pia mater, requiring participants to manipulate the instrument at extreme angles.
Such changes in orientation frequently moved the tracking markers out of the camera's field of view, resulting in higher rates of tracking loss. 
In addition, scissors were used far less frequently, typically only at the beginning of the procedure to make the initial incision in the pia. 
As a result, less tracking data were available for this instrument.

The level of tracking accuracy achieved by the system raises an important question about its adequacy in assessing surgical skill in a broader evaluative context.
While this question lies beyond the scope of the present study, it is reasonable to assume that more complete tracking data could enhance the ability to distinguish between levels of expertise in more complex surgical tasks.
Unlike virtual reality-based studies\cite{Yilmaz2022,Natheir2023}, which inherently provide complete and lossless motion data within a synthetic environment, physical simulations with optical tracking systems are subject to real-world limitations such as occlusion, marker orientation, and camera coverage. 
As a result, tracking loss is a challenge unique to physical simulation platforms and is generally not encountered in studies relying solely on virtual simulations. 

Another limitation is related to the small sample size, particularly within the Senior and Expert groups.
This constraint may have reduced the analyses' statistical power, possibly explaining the lack of significant differences between these groups.
Access to expert neurosurgeons for simulation studies is challenging due to limited time and clinical duties.
This may limit the scalability of future studies that involve high-expertise participants.
In addition, classifying participants by study year or expertise may not reflect their specific experience with the procedure.
For example, epilepsy fellows are more likely to be familiar with subpial resections than other subspecialities, which could contribute to increased within-group variability and blur the distinction between formal training levels.

Regarding the selected metrics, the motion-based metrics employed in the analysis were computed as global values over the duration of the trial.
While these metrics provide a high-level overview of instrument dynamics, they do not capture the temporal evolution of performance throughout the procedure. 
A more granular, time-resolved analysis could offer deeper insights into task segmentation, learning curves, and moment-to-moment variations in skill execution \cite{Forestier2012}.

Time-based metrics, while informative, cannot be interpreted in isolation as indicators of performance. 
A shorter procedure does not necessarily reflect greater proficiency, as it can come at the expense of accuracy, care, or safety.
Therefore, time should be interpreted in conjunction with qualitative and task-specific performance indicators.

This ex-vivo calf brain simulation platform provides a practical and anatomically realistic setting but does not encompass the full range of surgical competencies involved in the dynamic interaction between the  educator and the learner in the complex human operating room environment.
The absence of bleeding, vascular responsiveness, and tissue pulsation, normally associated with cardiac and respiratory activity, limits the model's ability to fully replicate intraoperative challenges. 
These factors are essential in influencing surgical decision-making and instrument handling under realistic conditions. 
Although post-mortem degradation typically affects tissue consistency and elasticity, in this study, care was taken to use fresh brain specimens, thereby minimizing such degradation and preserving tactile realism during the simulations. 
Moreover, anatomical differences between calf and human brains, including variations in cortical thickness, sulcal morphology, and vascular layout, may have affected performance of experts more than novice participants.

\section{Conclusion}
\label{sec:conclusion}

We developed and evaluated a neurosurgical simulation platform based on an ex-vivo calf brain model, designed to support the training and assessment of bimanual subpial corticectomy procedures. 
The platform integrates real-time instrument tracking with video-based annotation, enabling the extraction of objective kinematic and bimanual coordination metrics. 
We conducted a case series involving 47 participants across four training levels: medical students, junior residents, senior residents, and neurosurgeons.
On average, 81\% of the instrument usage time was successfully captured, involving the bipolar forceps, ultrasonic aspirator, and micro scissors.
We successfully derived objective metrics that demonstrated the platform’s ability to capture performance differences between varying levels of surgical expertise.
The results obtained highlight the potential of this platform as a basis for further studies to advance our understanding of the composites of expert surgical technical skill performance. 
These investigations help provide pathways for our future work focused on using machine learning to develop and test intelligent tutoring systems capable of real time assessment of psychomotor performance and providing continuously verbal automated trainee feedback \cite{Yilmaz2022,Yilmaz2024,Giglio2025}.
The integration of these intelligent tutoring systems with expert human educators will provide a foundation for the creation of the "\textit{Intelligent Operating Room}" which will optimize the assessment and training the learners while mitigating surgical error.

\bibliographystyle{ieeetr}
\bibliography{refs} 

\end{document}